\documentclass[twocolumn]{aastex631}
\usepackage{lineno}
\shorttitle{Astro-COLIBRI}
\shortauthors{Reichherzer et al.}
\graphicspath{{./}{figures/}}

\begin{document}
\small{This is the \textbf{Accepted Manuscript} version of an article accepted for publication in the \textbf{Astrophysical Journal Supplement Series}. IOP Publishing Ltd is not responsible for any errors or omissions in this version of the manuscript or any version derived from it. The Version of Record is available online at \textbf{10.3847/1538-4365/ac1517}.\\}
\title{Astro-COLIBRI --- The COincidence LIBrary for Real-time Inquiry for multi-messenger astrophysics}

\email{astro.colibri@gmail.com}

\author[0000-0003-4513-8241]{P. Reichherzer}
\affiliation{IRFU, CEA, Université Paris-Saclay, F-91191 Gif-sur-Yvette, France}
\affiliation{Ruhr-Universität Bochum, Universitätsstraße 150, D-44801 Bochum, Germany}
\affiliation{Ruhr Astroparticle and Plasma Physics Center, Ruhr-Universität Bochum, D-44780 Bochum, Germany}

\author[0000-0003-1500-6571]{F. Sch\"ussler}
\affiliation{IRFU, CEA, Université Paris-Saclay, F-91191 Gif-sur-Yvette, France}

\author[0000-0002-4806-8931]{V. Lefranc}
\affiliation{IRFU, CEA, Université Paris-Saclay, F-91191 Gif-sur-Yvette, France}

\author[0000-0002-6363-4284]{A. Yusafzai}
\affiliation{IRFU, CEA, Université Paris-Saclay, F-91191 Gif-sur-Yvette, France}
\affiliation{ECAP, FAU Erlangen-Nürnberg, D-91058 Erlangen, Germany}

\author[0000-0001-7964-4420]{A.K. Alkan}
\affiliation{IRFU, CEA, Université Paris-Saclay, F-91191 Gif-sur-Yvette, France}

\author[0000-0002-2153-1818]{H. Ashkar}
\affiliation{IRFU, CEA, Université Paris-Saclay, F-91191 Gif-sur-Yvette, France}

\author[0000-0002-1748-7367]{J. Becker Tjus}
\affiliation{Ruhr-Universität Bochum, Universitätsstraße 150, D-44801 Bochum, Germany}
\affiliation{Ruhr Astroparticle and Plasma Physics Center, Ruhr-Universität Bochum, D-44780 Bochum, Germany}



\begin{abstract}
Astro-COLIBRI is a novel tool that evaluates alerts of transient observations in real time, filters them by user-specified criteria, and puts them into their multiwavelength and multimessenger context. Through fast generation of an overview of persistent sources as well as transient events in the relevant phase space, Astro-COLIBRI contributes to an enhanced discovery potential of both serendipitous and follow-up observations of the transient sky.
The software's architecture comprises a Representational State Transfer Application Programming Interface, both a static and a real-time database, a cloud-based alert system, as well as a website and apps for iOS and Android as clients for users. The latter provide a graphical representation with a summary of the relevant data to allow for the fast identification of interesting phenomena along with an assessment of observing conditions at a large selection of observatories around the world.
\end{abstract}



\section{Introduction}
Astronomy and astrophysics are currently undergoing several fundamental changes, such as the increasing relevance of observations of transients, i.e. short-lived astrophysical phenomena such as supernova explosions, fast radio bursts (FRBs), and gamma-ray bursts (GRBs). At the same time, an increasing number of fundamentally new cosmic messengers provide crucial information about these objects. Today, the detection of high-energy neutrinos and gravitational waves (GWs) routinely supplement traditional astronomical observations in the electromagnetic spectrum \citep{Abbott_2017, Albert_2017, IceCube_170922A, Atel_10791}. These changes are driven by new instruments dedicated to the study of the most violent phenomena in the universe (e.g. supernovae, GRBs, and other bursts of extragalactic objects), as well as by improving real-time alert and follow-up systems \citep{4pisky_2016, AMPEL_2019, AMON_2020, Ashkar2021, FINK_2021}.

In the coming years, we will see the advent of a large variety of next-generation observatories dedicated to rapidly evolving time-domain astronomy and astrophysics. These observatories cover the full electromagnetic spectrum from the radio domain (e.g. Square Kilometre Array), optical observations (e.g. Large Synoptic Survey Telescope), X-rays (e.g. Space Variable Objects Monitor, ATHENA) to the highest energy gamma rays (e.g. Large High Altitude Air Shower Observatory, Cherenkov Telescope Array(CTA)). These are complemented by significant improvements and commissioning of observatories of novel messengers from the violent universe: high-energy neutrinos (e.g. IceCube, KM3NeT, Gigaton Volume Detector) and GWs (e.g. Virgo/LIGO/KAGRA, LISA). The related wealth of available information promises significant breakthroughs in various domains. On the other hand, the enormous increase of information that must be distributed globally, filtered, and classified in real time, etc., requires a fundamentally new way of data handling. Most of the decision chains reacting to the multiwavelength and multimessenger information must be fully automated. Rapid categorization of new information will allow for requesting crucial additional observations of the same event and thus helping to build a complete picture of the phenomena.

These new ways of carrying out astronomical observations have been accompanied by changes in communication between the scientists involved. The discovery of a new type of transient astrophysical phenomenon must be communicated very quickly within the community so that other observatories can make detailed follow-up observations. While parts of this information exchange use common, machine-readable data formats, e.g. those proposed by the International Virtual Observatory Alliance, the dissemination of subsequent observation results is still mostly communicated in human-readable form.

Astro-COLIBRI (COincidence LIBrary for Real-time Inquiry) tackles the challenges of the growing number of data and alerts via a fully automated alert classification and summarizing framework that will enable the astronomical community to quickly react to promising new phenomena or newly detected sources.

The paper is structured in a way that first explains the science drivers of Astro-COLIBRI in Section~\ref{sec:2}. Subsequently, in Section~\ref{sec:3} the modular architecture of this platform is explained and the high performance is emphasized in order to meet the demanding requirements in the context of real-time astronomy. The main functionalities are explained in Section~\ref{sec:4}, followed by the outlook in Section~\ref{sec:5}.

\section{Science drivers}\label{sec:2}
Currently, astrophysics is advancing from only studying stable celestial objects to also observing transients, i.e. short-lived astrophysical phenomena such as supernova explosions, GRBs, GWs, and flares of active galactic nuclei (AGNs). In the following, the different characteristics of the various transient events will be discussed, whose further investigation and improved follow-up observations are the scientific drivers of the Astro-COLIBRI software. The discussion of the different event types motivates the need for the flexible design of Astro-COLIBRI.

\subsection{Active Galactic Nuclei Flares}
Extreme conditions prevail in AGN jets due to strong magnetic fields, as well as relativistically accelerated plasmoids moving through ambient mass and photon fields, and generating secondaries such as neutrinos and photons of the entire electromagnetic spectrum through various mechanisms that include hadronic interaction processes \citep{Boettcher_2013, Hoerbe_2020}. An indication that AGNs are an important component of multimessenger astronomy has been provided by both the temporal and spatial co-relation of a blazar, a subclass of AGNs with jets extending toward Earth, with a neutrino measured by IceCube \citep{IceCube2018}. The emersion of AGN jets is characterized by a large variability on different time scales \citep{Hawkins_2007, Burd_2021, HESS2021A&A...648A..23H, 2021ApJ...911L..18K}, which requires rapid follow-up observation.\\

\subsection{Fast Radio Bursts}
FRBs are very short-flaring transients. The duration of the flares typically lasts between microseconds to milliseconds. The processes that produce these short, extreme flares are the subject of ongoing research. A first indication had been provided by the simultaneous detection of X-ray and radio bursts from the Galactic magnetar SGR 1935-243~\citep{2020ApJ...898L..29M}. Because of the short burst durations, follow-up observations of the afterglows are particularly important, since direct observation of the burst is typically not possible outside of coordinated multiwavelength campaigns.

\subsection{Gamma-Ray Bursts}
GRBs are powerful explosions that occur on relatively short time scales of a few milliseconds to a few hundred seconds. The sky is constantly monitored in the gamma-ray domain by Fermi and INTEGRAL, and in the X-ray domain by Swift, producing about one alert per day on average. The fundamental phenomena giving rise to the high fluence of GRBs are not fully understood. GRBs have now been firmly established as very high-energy (VHE) emitters~\citep{2019Natur.575..464A, Magic_2019, GRB190829A_HESS}. However, at the same time recent intriguing observational results raise questions on the dominant VHE production mechanism, its duration and evolution over time, the required properties of the progenitor as well as the circumburst environment, the maximum energy reached, and its possible scaling with the jet opening angle, etc. Continued rapid follow-up observations across the electromagnetic spectrum will be crucial to answering them. Additional input to this quest has also been provided by the first observational link established between mergers of binary neutron stars, short GRBs, and optical emissions known as kilonovae via the detection of electromagnetic signals following the GWs from the merger of a binary neutron star system~\citep{Abbott_2017}. 

\subsection{Gravitational Waves}
According to the general theory of relativity, GWs are waves in space-time that propagate at the speed of light. GWs are predicted to originate from particularly massive, strongly accelerated object systems, which include nonrotationally symmetric systems but also the merging or asymmetric collapse of massive objects, such as black holes or neutron stars. Advanced interferometer-based GW observatories, such as LIGO and Virgo, have successfully detected an increasing number of such events and have therefore added another component to the ensemble of particles used in multimessenger astronomy. This is opening up new, promising, and often complementary follow-up opportunities \citep{Ashkar2021}. Increasing interest in these efforts is providing a rich and rapidly increasing volume of observational reports. However, the search for multimessenger and electromagnetic counterparts and associated sources of GWs is proving very challenging due to large uncertainties in the sky localization of current GW observatories. Over the next few years, the Advanced Virgo and Advanced LIGO observatories will further increase their sensitivity and the KAGRA interferometer in Japan, as well as LIGO-India, will join the global network~\citep{2018LRR....21....3A}.

\subsection{High-energy Neutrinos}
\begin{figure}
\centering
\includegraphics[width=0.7\linewidth]{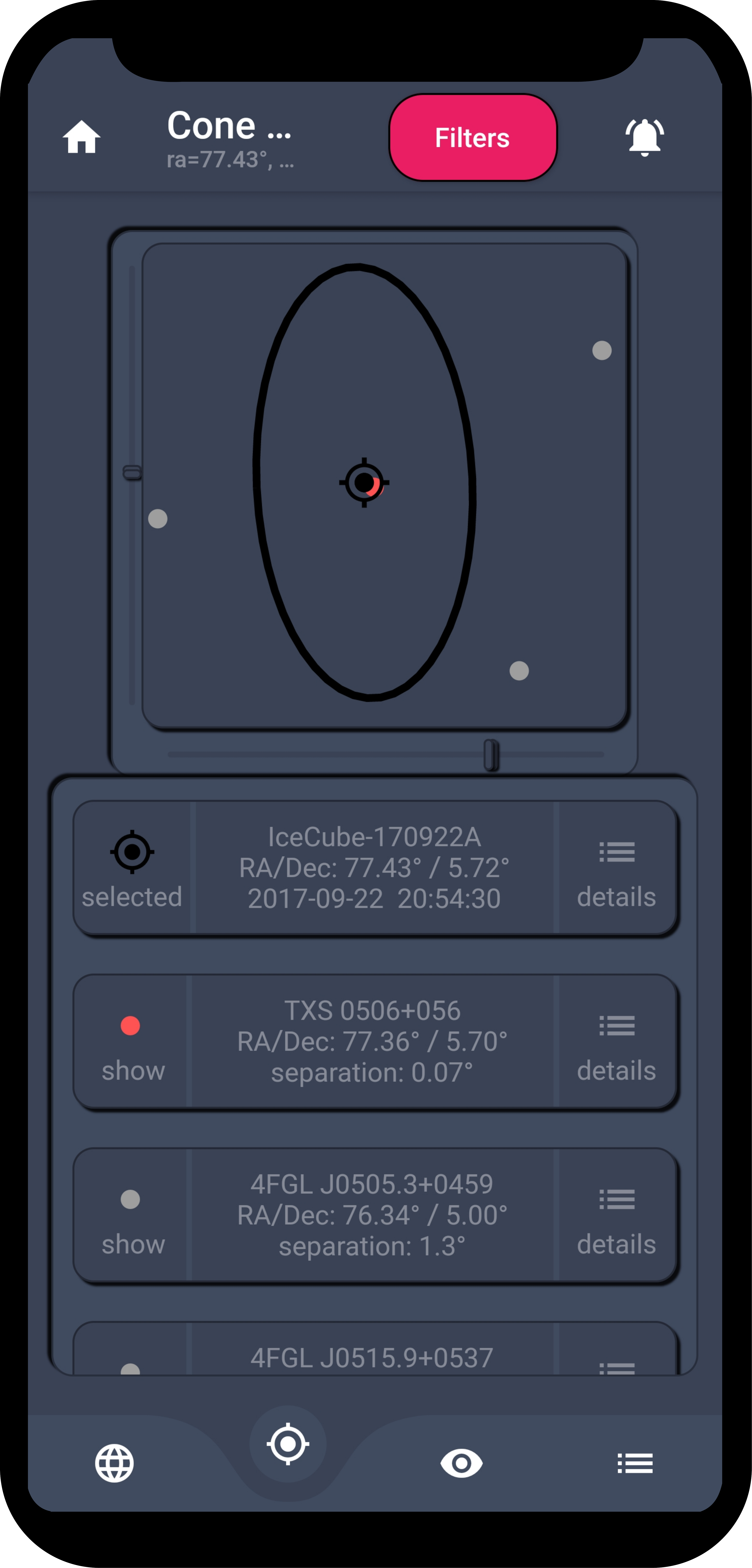}
\caption{Astro-COLIBRI mobile view of the cone search of the IceCube alert 170922A, which is marked by the black sign in the sky map. In the upper part, the IceCube event is shown in the context of known sources (known TeV sources are red -- others in gray) from catalogs (see Table~\ref{tab:catalogs}). The blazar TXS 0506+056 is close to the detected neutrino position within the uncertainty region, which is indicated by the black ellipse. Below the sky map, the sources are listed in ascending order of the separation of the event. Details and further links to each source and event are accessible through the details button. \label{fig:icecube}}
\end{figure}

Neutrinos reach us in an almost-unaffected state due to their low level of interaction, giving us a deeper insight into the central areas of astronomical particle accelerators. Relativistically accelerated cosmic rays produce high-energy astrophysical neutrinos in interactions with the surrounding matter or photons in these accelerators \citep{2008PhR...458..173B, BeckerTjus2020}. Observations of astrophysical neutrinos have increased in recent years. The current generation of high-energy neutrino telescopes like IceCube and ANTARES report on detections that are rarely able to be associated with astrophysical sources.
With the blazar TXS 0506+056, only one potential extragalactic source of neutrinos has been identified to date. Figure~\ref{fig:icecube} shows this neutrino detection location, as reported by IceCube via the IceCube-170922A alert~\citep{IceCube_170922A}. The blazar TXS 0506+056 can easily be seen to lie well within the uncertainty region of the neutrino event. Unfortunately, this identification of TXS 0506+056, as well as the detection of its flaring state, had taken several days~\citep{Atel_10791}, a latency that is not commensurate with real-time astronomy. This example illustrates the difficulties in finding additional point sources of high-energy neutrinos and highlights the importance of an easy-to-use interface with graphical assistance in finding them.\\\\

\subsection{Tidal Disruption Events (TDEs)}
A TDE typically results from an orbiting star approaching its hosting supermassive black hole (SMBH) closely enough to be disrupted by the gravitational forces. Recently published studies \citep{2021NatAs...5..510S, Winter2021} highlight the importance of these events for transient astronomy, through the very bright, long-lasting flares that extend over a broad wavelength spectrum. The possible association of neutrinos with TDEs~\citep{2021NatAs...5..510S} triggered renewed interest in this research direction, to be confirmed by future observations.

\section{Architecture / design}\label{sec:3}
\begin{figure*}
\centering
\includegraphics[width=1.0\linewidth]{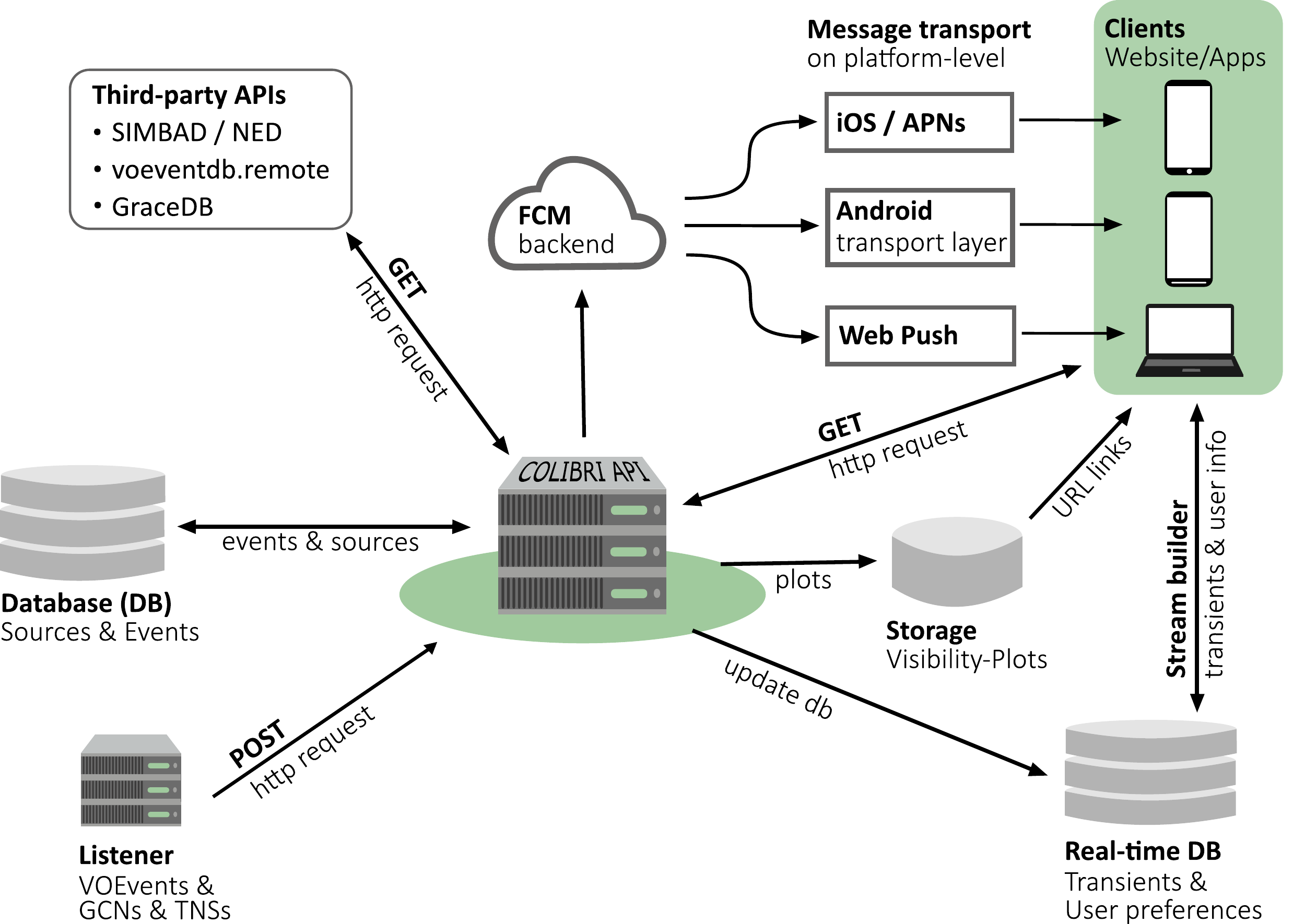}
\caption{Architecture of Astro-COLIBRI. The individual modules are described in detail in the text. The fields highlighted in green are public interfaces.  \label{fig:architecture}}
\end{figure*}
In multimessenger astronomy, to tackle the above-described science cases, among others, an ecosystem of numerous complimentary services has been established over the last decades. Astro-COLIBRI aims to be the top layer that connects existing subsystems into a large ecosystem, with a focus on optimized display for users through an interactive graphical user interface (GUI) and mobile apps that receive push notifications in real time, both usable in an intuitive way. In doing so, Astro-COLIBRI uses existing alert and correlation systems such as VOEvent alerts, Astrophysical Multimessenger Observatory Network alerts \citep{AMON_2020}, Transient Name Server (TNS) notifications\footnote{\url{https://www.wis-tns.org/}}, Gamma-ray Coordinates Network (GCN) circulars\footnote{\url{https://gcn.gsfc.nasa.gov/}}, brokers (e.g. via 4 Pi Sky; \citet{4pisky_2016}, FINK broker; \citet{FINK_2021}), and VOVisObs protocols for scheduled observations. 

To accommodate the growing range of services from the community, Astro-COLIBRI is modular in design and allows easy integration of new or modifying third-party services. Details of the modular design are presented in the following subsections, which also specify the communication standards that Astro-COLIBRI uses to enable optimized communication both within the Astro-COLIBRI system and to the integrated and linked third-party providers within the astronomical community. Figure \ref{fig:architecture} presents an overview of the Astro-COLIBRI architecture that will be presented in this section in detail.

Overall, the various modules are deployed in such a way that the technologies used in them optimize the performance of the entire alert pipeline as well as the general usability of the system in terms of speed, an aspect that has the highest priority in real-time multimessenger astronomy. Table \ref{tab:performance} provides an overview of the performance of individual tasks within Astro-COLIBRI. Harder to measure directly, additional drivers for the development of the system have been their accessibility and ease of use.
\begin{table}
    \centering
	\caption{Performance Overview of Key Functionalities of Astro-COLIBRI Discussed in Sections~\ref{sec:3} and \ref{sec:4}.}\label{tab:performance}
	\begin{tabular}{ccc}
		\hline 
		\hline
		Functionality & Time (ms) & Comment \tabularnewline
		\hline 
		Push notification alert & 335\footnote{Taken from \citet{ALBERTENG2020}} & FCM pipeline \tabularnewline
		Cone search of transient & $1291 \pm 71$ & Firestore query \tabularnewline
		Manual cone search & $850 \pm 292$ & API request \tabularnewline
		Monthly visibility plot & $892 \pm 59$ & API request \tabularnewline
		Detailed visibility plot & $2768 \pm 80$ & API request \tabularnewline
		\hline 
	\end{tabular}
	{\raggedright \textbf{Note.} The presented time is the average time needed to perform the action. \par}
\end{table}

\subsection{RESTful API}
Our central point in the overall architecture is the Representational State Transfer (REST or RESTful) Application Programming Interface (API) built within the Flask framework using Python as the programming language, which allows the utilization of many well-established astronomy packages such as astropy~\citep{2018AJ....156..123A} and astroquery~\citep{Astroquery_2019}. Clients and external services can send HTML requests to the API through their endpoints. The API performs the specified tasks and returns a JSON file with the computed and collected information. Our API runs in a container and can thus be flexibly deployed on any cloud-computing platform. Currently, the API is hosted on Heroku\footnote{\url{https://www.heroku.com/}}.\\\\
The Astro-COLIBRI API uses the model-view-controller pattern with the following structure:
\subsubsection{Models}
Models get called by the controllers, handle all computations, and are responsible for retrieving data from third-party services and our own database. Models also process and store the data locally such that it can be sent directly by the controller back to the clients.
\subsubsection{Views}
The views contain all HTML and CSS files that display the static content of the API website. We intend to only document the API usage here and provide examples.
\subsubsection{Controller}
The controllers are responsible for understanding the client requests and then initiating the correct calculations by calling the model functions.

Via Python, the API has many access points to important third-party services and is therefore used for all tasks involving these services. All endpoints modifying our database are password protected and only accessible to our collaborators. The main task using private endpoints is updating, modifying, or merging transient events in our database based on public alerts we listen to or private alerts from our collaborators.\\\\
The majority of our endpoints are publicly accessible and are thus used by the community and the Astro-COLIBRI clients (see our documentation\footnote{\url{https://astro-colibri.herokuapp.com/documentation}} for technical details). The public endpoints include the following services.
\subsubsection{Cone Search} 
Cone searches are one of the main functions in Astro-COLIBRI (see Section \ref{sec:cone_search}) and allow the representation of an event or a source in the context of other already known events or sources in the relevant temporal and spatial phase space.
\subsubsection{Visibility Plot} 
The API provides a graphical representation of the visibility of an arbitrary sky position for the next 24 hr via this endpoint, customized to the observatory relevant for the user. Figure~\ref{fig:vis} shows an example of a visibility plot for the H.E.S.S. Observatory~\citep{Aharonian:2006pe}, where the observatory's specific details, such as observations during partial moonlight, are taken into account. The detailed, daily plot is complemented by a monthly overview plot of the observability in the observatory specified by the user.

\subsection{Clients}
To notify users in the most effective way and to provide modern and mobile usage we decided to develop platform-independent Astro-COLIBRI user interfaces. Using the open-source framework Flutter, we developed a website and apps for iOS and Android natively in the programming language dart. Since we only use one common codebase for all clients, the maintenance of the code remains effortless and the implementation of further features particularly fast.
The website \url{www.astro-colibri.com} is hosted in Firebase and is accessible via compressed files cached on solid-state drives at content delivery network edges worldwide provided by Google allowing for performant, location-independent access to the website.

The clients are connected to our real-time database via streams. For the real-time database, we use Firebase Firestore\footnote{\url{https://firebase.google.com/docs/firestore}}. Transient events which are interesting for the user are displayed via these streams in real time and announced via push notifications on mobile devices. More computer-intensive or third-party operations are outsourced from the front end to the Astro-COLIBRI API.

\subsection{Event Listener}
The event listener permanently runs as a background process on a dedicated server that has the capacity to listen to VOEvents, TNS notifications, and GCNs. While the VOEvent listener is building upon the {\it Comet} broker~\citep{2014A&C.....7...12S}, the analysis of TNS notifications and GCN circulars is relying on a custom-build, Python-based email parser. For the science cases defined in Astro-COLIBRI, relevant notifications are received by our listener and passed to the Astro-COLIBRI API via post requests for further processing. The API checks whether it is a new event or an update with more accurate information on an existing event, and merges the available data by also complementing the data with third-party (e.g. TNS API) information. \\

\subsection{Databases}
Due to the flexible structure and the number of different event types, we decided to use a non-Structured Query Language data structure, where data is stored in documents, organized in collections. Individual documents can contain complex nested objects in addition to subcollections allowing for a nonprescriptive choice of parameters for individual documents. In order to meet the demanding functionalities of Astro-COLIBRI, two databases are used that are optimized for different tasks.

The main storage of events relevant for Astro-COLIBRI is done in a Mongo-DB which is continuously updated via the API by new VOEvents or other notifications as well as manual requests. Only the API has access to this database and uses its information for the processing of new events and the execution of user requests, such as the cone searches.

The Firebase Firestore real-time database, on the other hand, provides a direct stream to all clients and allows the information to be displayed in real time on the front end devices, without the need to reload the web interface or application. 
This database is populated directly from the API and is automatically kept up to date. The exchange between the front end devices and the real-time database is characterized by the following features:
\subsubsection{Offline Support} 
All relevant data is saved locally on devices and all changes are automatically synchronized between the clients and the database when the device comes back online.
\subsubsection{Expressive Querying}
The support of chained and combined filters and queries in combination with the usage of an efficient zig-zag merge join algorithm offers both high-performance and flexible operation.
\subsubsection{Real-time Streams}
Usage of data synchronization via streams allows for updating the displayed information on any connected device within (milli)seconds (see Table~\ref{tab:performance}).

\subsection{File Storage}
Files created in Astro-COLIBRI are stored in Firebase storage and automatically managed and eventually deleted once not needed anymore. For example, the visibility plots are stored here, so that only a link to the storage location is sent to the end devices, via which the desired plots can be displayed to the users.
The plots can also be easily downloaded by the user via the clients.

\subsection{Message Transport System}
Astro-COLIBRI sends cross platform push notifications via the cloud-based Firebase Cloud Messaging (FCM) service\footnote{\url{https://firebase.google.com/docs/cloud-messaging}}. 
Users can personalize in their Astro-COLIBRI preference settings which events they want to be notified about via push notifications. If there is interest in a certain event class, the user's device will be subscribed to the dedicated topic. The Astro-COLIBRI API sends push notifications to all devices that subscribed to the corresponding topic of a new transient event.

\section{Functionalities}\label{sec:4}

\subsection{Real-time alert pipeline for transient events}
In real-time, multimessenger astronomy, notifications of transients and the assessment of their relevance to the respective observatories for follow-up observations are, by their very nature, of fundamental importance. There are numerous channels in multimessenger, multiwavelength astronomy for this relevant information on both different platforms and different time scales. 
\begin{figure}
\centering
\includegraphics[width=1.0\linewidth]{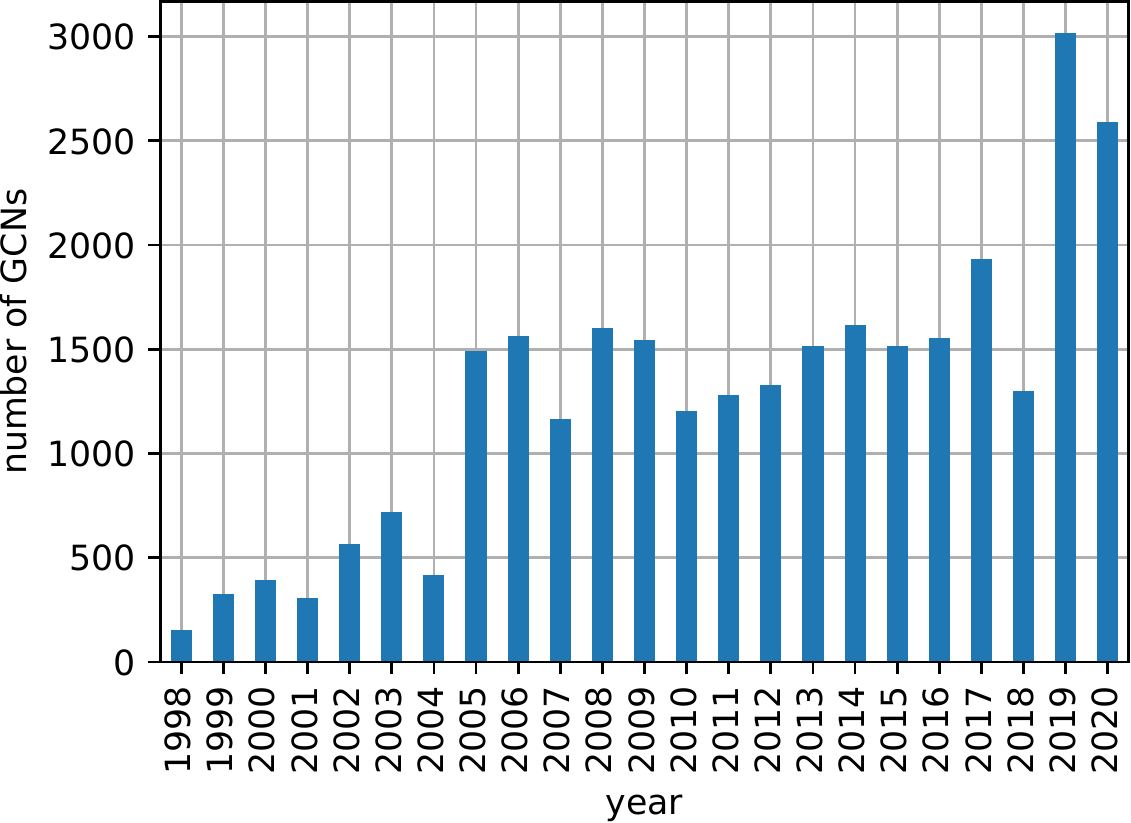}
\caption{The number of GCN circulars has steadily increased over the recent years. \label{fig:gcn_statistics}}
\end{figure}
For example, there are various alert systems, such as VOEvents, GCNs (notices and circulars), TNS notifications, and The Astronomer's Telegram (ATEL), which differ both in their degree of machine or human-generated content and in the underlying communication technology. The technologies involve the use of web pages, emails, and brokers and require the end user to make an effort to keep track of (the ever-increasing amount of) the relevant information. In addition, it should be noted that several observatories may be detecting the same astrophysical event and that several alerts for the same event are emitted by each observatory. The challenge for the community is to keep track of this information in real time to enable informed decisions on additional observations.

The increasing number of GCNs is demonstrated in Fig.~\ref{fig:gcn_statistics}. With new observatories like the Vera Rubin Observatory, CTA, and KM3NeT coming online in the next few years, the flood of information will keep growing strongly. Astro-COLIBRI takes a cross platform approach and processes the alerts of the different platforms automatically in real time and creates a top layer representation of the transient events in multimessenger astronomy by a correlated examination and underincorporation of further historical data easily visible for the user.

\begin{figure}
\centering
  \centering
  \includegraphics[width=1.0\linewidth]{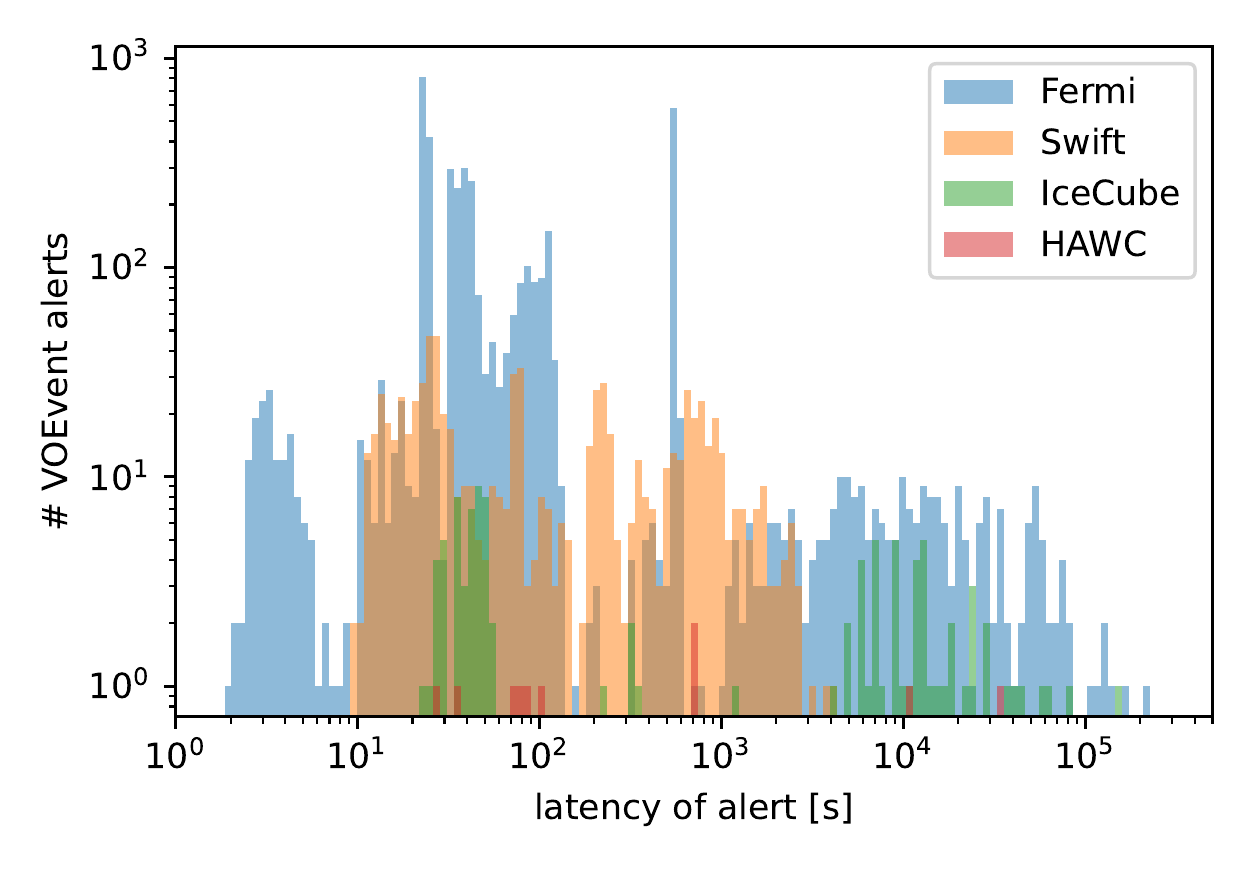}
  \caption{Latency of the distribution of VOEvents from different observatories during the window between 2018 January 1 and 2021 June 22. We show VOEvents from the Fermi Gamma-ray Burst Monitor, Swift Burst Alert Telescope, Swift X-ray Telescope, and VOEvents from the High Altitude Water Cherenkov burst monitor. All VOEvents whose ivorn contain \textit{amon/icecube}, \textit{ICECUBE\_GOLD}, \textit{ICECUBE\_BRONZE}, \textit{ICECUBE\_Cascade\_Event}, or \textit{NU\_EM\_Event} are summarized in IceCube.}
  \label{fig:statistics1}
\end{figure}

Astro-COLIBRI constantly and automatically scans reports on transient activity and processes new or updating alerts in a real-time pipeline, which is built up from the modules described in the following.
The pipeline is triggered when the Astro-COLIBRI alert listener receives a VOEvent alert via one of the public VOEvent brokers\footnote{\url{https://wiki.ivoa.net/twiki/bin/view/IVOA/IvoaVOEvent}}. For redundancy, Astro-COLIBRI is subscribed to at least three brokers simultaneously. Additional alerts are received by notifications from the TNS. An automated treatment of GCN circulars provides additional context like the names assigned to transient events. If the received information matches the science goals defined in Section~\ref{sec:2}, the listener sends the information to the Astro-COLIBRI API for further processing. Taking into account the services described in Section~\ref{sec:3rd_party_services} and listed in Table~\ref{table:3rd_party_services}, the event described in the alert is supplemented with data and information, or at least complemented with links to various external services, and pushed to the Astro-COLIBRI databases. Here, the question of whether it is a new event or an update of a previously stored event is decided, which results in a merge process of the available information. Figure~\ref{fig:statistics1} shows the number of VOEvents as a function of the latency between the alerts being sent with respect to the exact event time. Astro-COLIBRI handles the various updated information sent with alerts by this dedicated pipeline.
\begin{figure*}
\centering
\includegraphics[width=1.0\linewidth]{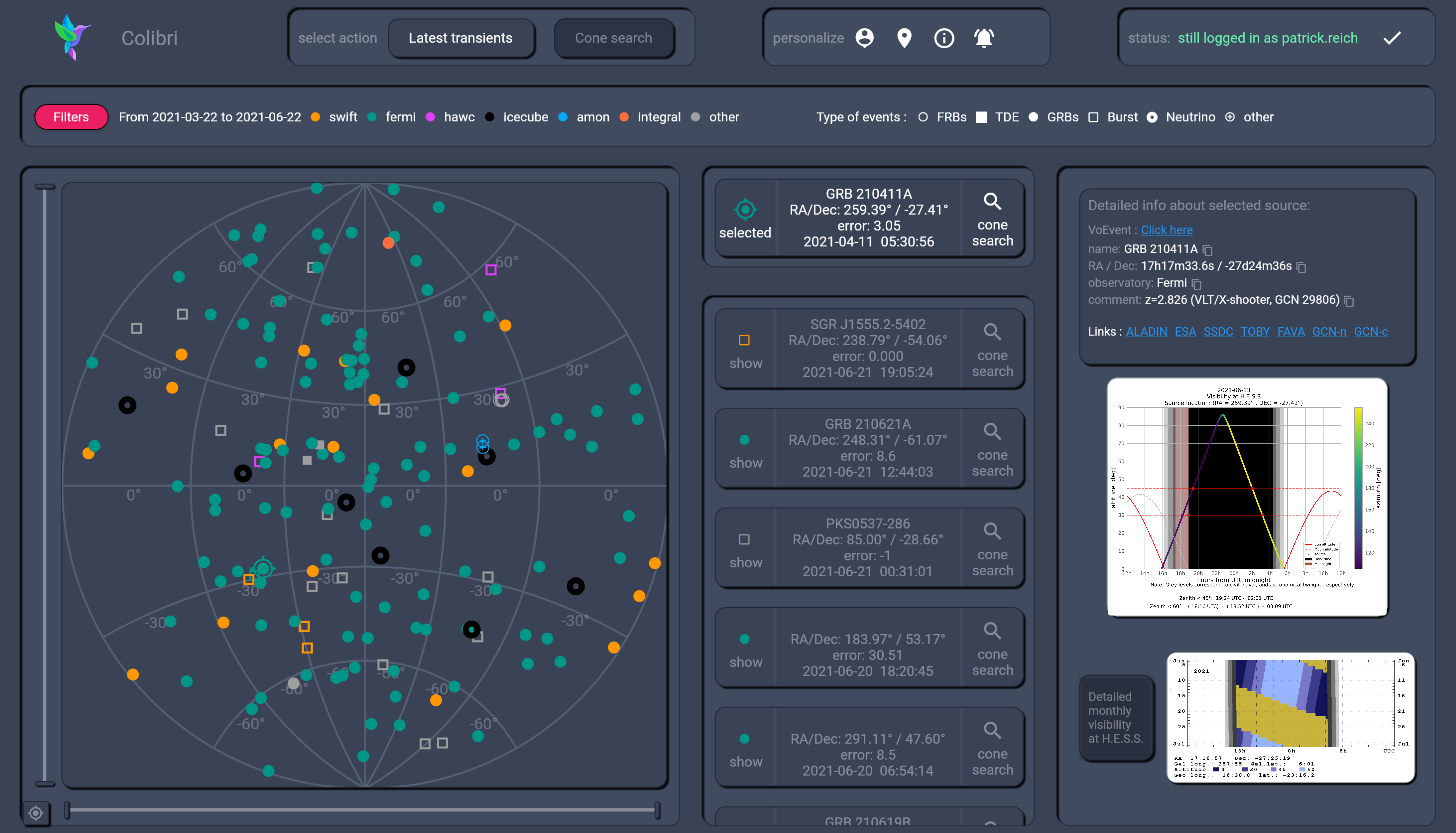}
\caption{Astro-COLIBRI overview of latest transients. The top area provides menus for filtering and setting up user preferences. The main information panel below is divided into three subareas: the sky map, the list of events and sources, and the final panel with further details, links and visibility information. Screenshot taken on 2021 June 21. \label{fig:latest_transients}}
\end{figure*}
The notifications are sent via the FCM protocol to all mobile devices with the Astro-COLIBRI app that have indicated an interest in this event type via their personal settings. In parallel, the update in the real-time database leads to an update in the GUI of the website and apps, through the direct connection via streams. For example, the sky map is automatic including all transient events according to the user-specific filter criteria (see Section \ref{sec:filters}). Figure~\ref{fig:latest_transients} presents the website, where an overview of the latest transients is given in the sky map and the list of events. Detailed information and links to relevant services are given for the selected source or event in the right panel.

\subsection{Cone Searches}\label{sec:cone_search}
Besides the main functionality to notify users about transient alerts in real time, Astro-COLIBRI presents the events in the context of historical data, as well as other transient events under consideration of time scales relevant for the respective event category. For this purpose Astro-COLIBRI uses cone searches around the event of interest, in which sources and transient events are shown, the latter being filtered for on the user-adjustable relevant time scale. An example cone search is presented in the website view in Figure~\ref{fig:icecube}. The default radius of the cone searches is 10° and is visualized by the gray boundary (not shown here due to being zoomed in by a few degrees). The uncertainty of the observation is visualized by the boundary matching the color of the marker. In this example, it is the black ellipse.

The sky map, in which the events and sources are shown in the relevant phase space, can be zoomed and panned. Sources and events can be selected directly from the map or from the attached list. In the list, the sources and events contained in the cone search are listed in ascending order according to the distance to the cone-search center. Table \ref{tab:catalogs} lists the catalogs currently used in the Astro-COLIBRI cone searches. 

For the case of GWs, contours of different confidence levels are shown instead.\\\\

\begin{table}
    \centering
	\caption{Sources available in Astro-COLIBRI Cone Searches}\label{tab:catalogs}
	\begin{tabular}{ccc}
		\hline 
		\hline
		Catalog name & Number of sources & Citation \tabularnewline
		\hline 
		Fermi 4FGL & 5788 & \citet{4FGL2020} \tabularnewline
		TeVCat & 118 & TeVCat catalog\footnote{\url{http://tevcat.uchicago.edu/}} \tabularnewline
		FlaapLUC & 228 & \citet{FlaapLUC2018} \tabularnewline
		\hline 
	\end{tabular}
	{\raggedright \textbf{Note.} Astro-COLIBRI stores a merged catalog that combines all relevant information of the listed catalogs. The number of sources used in Astro-COLIBRI is also listed, as of 2021 May 21. This will be updated continuosly. \par}

\end{table}

\subsection{Interface with the Multimessenger Ecosystem}\label{sec:3rd_party_services}
\begin{figure}
\centering
\includegraphics[width=1.0\linewidth]{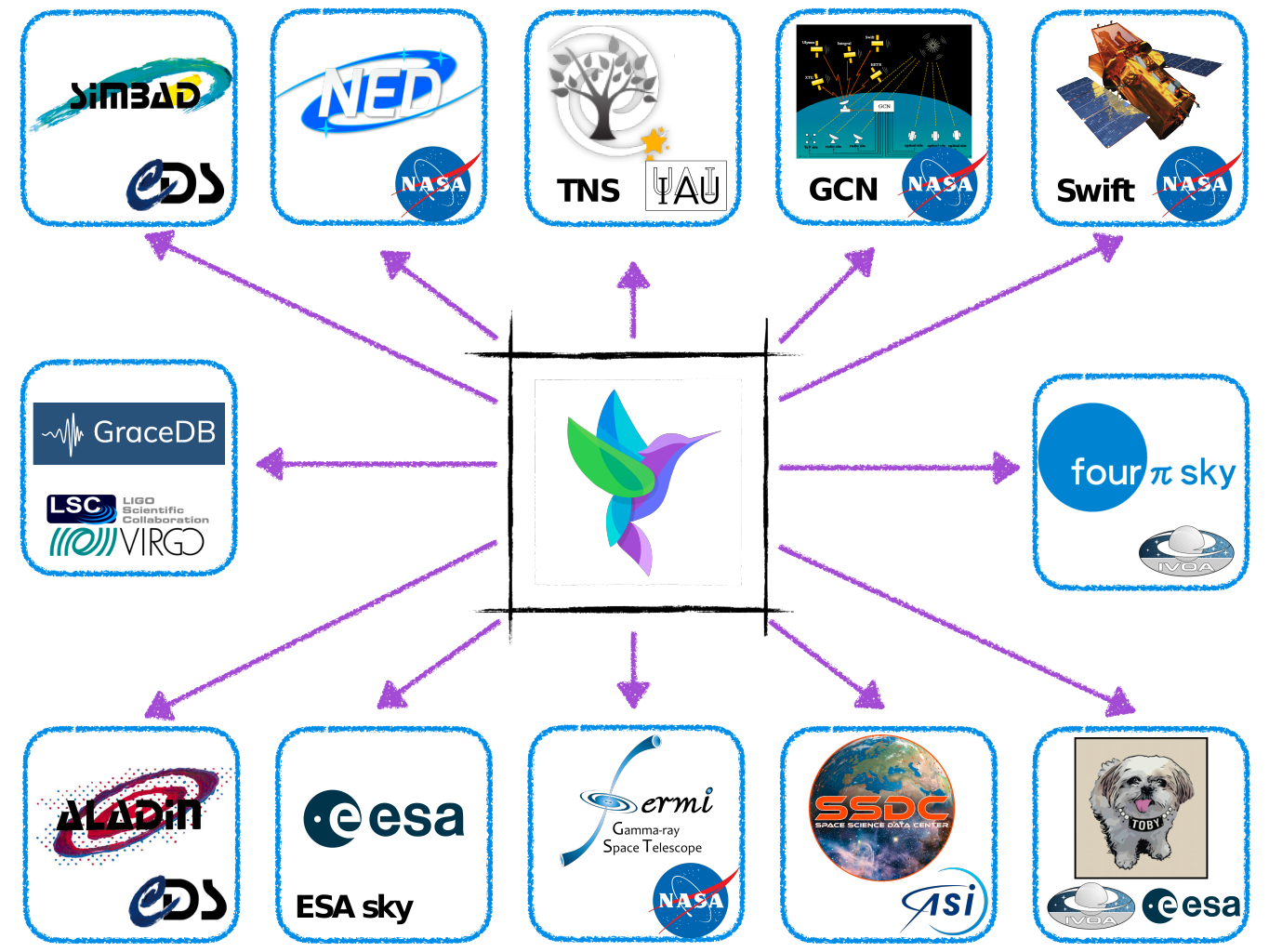}
\caption{Astro-COLIBRI provides direct links to dedicated services and websites. See Table~\ref{table:3rd_party_services} for details \label{fig:services}}
\end{figure}

A key feature is the listing of relevant information, especially services and information offered from the astronomical community for the user-selected event or source. For this purpose, a separate area is available for the clients. For each event and source selected in Astro-COLIBRI, customized links to a variety of external services are generated. This allows for direct access to supplementary information and allows for the use of Astro-COLIBRI as a single platform providing the user with an easy-to-access overview of relevant information. An overview of the currently used external services is provided in Figure~\ref{fig:services} and Table~\ref{table:3rd_party_services}.
\begin{table*}
    \centering
	\caption{Astro-COLIBRI provides customized links to the following services for each event and source (if available) to provide convenient access to specific, more detailed information on demand.}\label{table:3rd_party_services}
	\begin{tabular}{ccc}
		\hline 
		\hline
		Service & Description & Reference/Link \tabularnewline
		\hline 
		ALADIN & \multicolumn{1}{p{12cm}}{Displays sources from astronomical catalogs or databases in a sky region} & \citet{Aladin_2014}  \tabularnewline
		BAT & \multicolumn{1}{p{12cm}}{BAT GRB event data-processing report includes detailed information, a flux summary, spectra, and light curves with varoious binning} & \citet{BAT_2013} \tabularnewline
		FAVA & \multicolumn{1}{p{12cm}}{Fermi All-sky
        Variability Analysis (FAVA) for the selected position observed by the Large Area Telescope (LAT)} & \citet{FAVA_2017} \tabularnewline
		GCN-c & \multicolumn{1}{p{12cm}}{GCN circulars (GCN-c) deliver information in human-written reports about the observed event} &  GCN archive\footnote{\url{https://gcn.gsfc.nasa.gov/gcn/gcn3_archive.html}} \tabularnewline
		GCN-n & \multicolumn{1}{p{12cm}}{GCN notices (GCN-n) deliver information about basic properties of transient objects in automatically generated alerts} &  GCN website\footnote{\url{https://gcn.gsfc.nasa.gov/}} \tabularnewline
		NED &  \multicolumn{1}{p{12cm}}{NASA/IPAC Extragalactic Database (NED) concerning galaxies and other extragalactic objects, with more details, photometry, spectra, and further references/links} & \citet{NED_1991} \tabularnewline
		SIMBAD &  \multicolumn{1}{p{12cm}}{Provides additional information, cross identifications, bibliography, and measurements of sources} & \citet{Simbad_2000} \tabularnewline
		SSDC & \multicolumn{1}{p{12cm}}{Display of spectral energy distributions (SEDs) of astrophysical sources from the Space Science Data Center (SSDC)} & SED builder\footnote{\url{https://tools.ssdc.asi.it/SED/}} \tabularnewline
		Swift &  \multicolumn{1}{p{12cm}}{High-level information on the Swift BAT or XRT observations and links to not only these, but also other relevant references} & Swift GRB table\footnote{\url{https://swift.gsfc.nasa.gov/archive/grb_table/}} \tabularnewline
		TeVCat & \multicolumn{1}{p{12cm}}{Source catalog for VHE gamma-ray astronomy with detailed source information and linked references} & \citet{TevCat2008} \tabularnewline
		TOBY & \multicolumn{1}{p{12cm}}{The Tool for Observation visiBilitY and schedule (TOBY) shows the event visibility and schedule for a variety of observatories following the ObjVisSAP and ObsLocTAP protocols} & TOBY website\footnote{\url{http://integral.esa.int/toby/}} \tabularnewline
		XRT & \multicolumn{1}{p{12cm}}{Light curve, spectra, and comparison with other bursts provided by Swift's XRT}  & \citet{XRT_2009} \tabularnewline
		\hline 
	\end{tabular}
	{\raggedright \textbf{Notes.}\par}
\end{table*}

\subsection{Visibility of Sources for Observatories}
\begin{figure}
\centering
\includegraphics[width=1.0\linewidth]{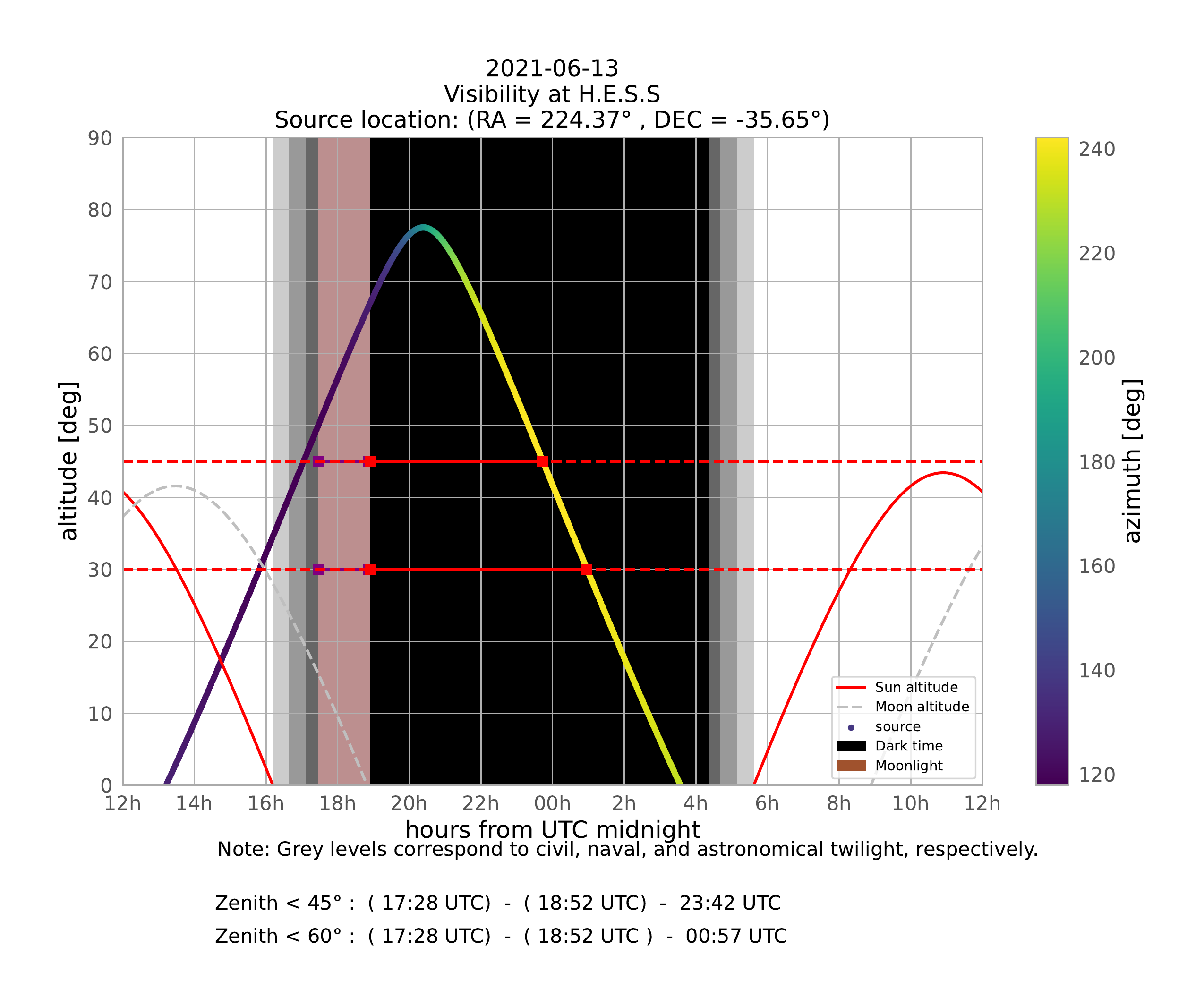}
\caption{Detailed visibility plot for the next 24h for the selected observatory H.E.S.S. of PKS~1454-354.\label{fig:vis}}
\end{figure}

In multiwavelength and multimessenger astronomy, there is an increasing demand for multimission coordination for follow-up observations of transient events. Astro-COLIBRI shows the near-term and long-term visibility of transient events of some predefined observatories or custom locations. The visibility plots contain the monitoring of source altitude, Sun and Moon altitude, Moon phase, Moon-to-source separation if available, etc. In the website and apps, the visibility for the user-selected observatory is shown immediately for the events/source selected by the user. The exact observation conditions, including Moon time, are implemented for selected observatories already. Figure \ref{fig:vis} shows an example of a visibility plot for H.E.S.S. that contains observatory-specific details. Further observatories will be included in the future. Via a customized link to the Tool for Observation visiBilitY and schedule (TOBY), users can see visibility details for the observatories INTEGRAL, Gaia, Chandra, Insight-HXMT, XMM-Newton, Nordic Optical Telescope, and Swift with one click in Astro-COLIBRI for the interesting source or event. 

\subsection{Event and Source Filter}\label{sec:filters}
To cover the main science goals specified in Section \ref{sec:2}, Astro-COLIBRI accommodates a variety of different event types, astrophysical messengers, wavelengths, and observation types. A filter menu allows users to select the most relevant one for their own interests. Three hierarchical levels of filtering are implemented. The first one is the wavelength and this has an impact on the two levels below: observatories (Fermi, IceCube...) and types of events (GRBs, neutrinos, TDEs...). Selecting observatories impacts only the type of event filter, and the type of event filter has no impact on the two above. In the sky map and the listing of all events, colored icons indicate the event type and detecting observatories, as shown in the legend.

\subsection{User management and access of private collaboration information}
In Astro-COLIBRI users can create an account, with the advantage that preferences, such as the possible filters from Section \ref{sec:filters}, or the selection of a preferred observatory, and therefore the customization of the visibility plot can be saved and used across all platforms. In addition, users can request access to proprietary information and private alerts of observatories upon proof of being affiliated with that observatory. For example, private FlaapLUC~\citep{FlaapLUC2018} based alerts on Fermi-LAT detected flares are only accessible to members of the H.E.S.S. collaboration. Further agreements of this kind are planned. 

\section{Outlook}\label{sec:5}
Astro-COLIBRI now offers a real-time overview of the transient sky and provides a high-level overview of relevant information about events. As a holistic system with a user-friendly graphical web interface that acts as a top layer of many specific services and information streams, it is suitable for use in observatory control rooms or as a mobile version on the cell phones of shifters or burst advocates. Astro-COLIBRI constantly listens to machine-readable information (VOEvent messages) about transient events propagated within the multimessenger community and informs users based on their preferences about these events in real time via push notifications on smartphones and notifications on the website. Astro-COLIBRI enhances this functionality with the ability to send private alerts from collaborating observatories to authorized users. The results of dedicated analysis of Fermi-LAT data via FLaapLUC alerts have already been integrated into the Astro-COLIBRI alert pipeline and provide real-time alerts to verified H.E.S.S. users for H.E.S.S. observations of eligible flaring AGNs. Further partnerships of this kind are planned for the future. Concrete discussions with other collaborations are already ongoing. 

Although the machine-readable alerts already contain information necessary for observations, additional detailed human-written reports (GCNs, ATEL) are circulated in the multimessenger community. Since these reports contain further relevant information, which may be crucial for the decision of follow-up observations, we will implement the functionality to process this information and enhance the services provided by Astro-COLIBRI with the supplementary information. This long-term project is based on natural language processing techniques, which evaluate these circulated human-written texts and reflect them in Astro-COLIBRI.

The Astro-COLIBRI development team welcomes comments and feedback from the community to further improve the platform and can be contacted at \href{mailto:astro.colibri@gmail.com}{astro.colibri@gmail.com}\\


This work was supported by the European Union’s Horizon 2020 Programme under the AHEAD2020 project (grant agreement No. 871158). This work is supported by the ``ADI 2019’’ project funded by the IDEX Paris-Saclay, ANR-11-IDEX-0003-02 (P.R.). P.R. also gratefully acknowledges support from the German Academic Exchange Service and by the RUB Research School via the \textit{Project International} funding.\\

\software{4 Pi Sky~\citep{4pisky_2016}, Astropy \citep{2013A&A...558A..33A,2018AJ....156..123A}, Astroquery~\citep{Astroquery_2019}, Flutter\footnote{\url{https://flutter.dev/}}}

%

\vspace{5mm}

{{\large \textit{Code availability:}}}~~~The code for the Astro-COLIBRI API, the front end, as well as the results and data in this paper are available to interested researchers upon request.\\


\bibliography{bib}{}
\bibliographystyle{aasjournal}



\end{document}